\documentclass[12pt]{article}
\def\be{\begin{equation}}
\def\ee{\end{equation}}
\def\bea{\begin{eqnarray}}
\def\eea{\end{eqnarray}}
\usepackage{graphicx}% Include figure files

\catcode`\@=11
\def\lsim{\mathrel{\mathpalette\@versim<}}
\def\gsim{\mathrel{\mathpalette\@versim>}}
\def\@versim#1#2{\vcenter{\offinterlineskip
\ialign{$\m@th#1\hfil##\hfil$\crcr#2\crcr\sim\crcr } }}
\catcode`\@=12
\usepackage{axodraw}

\catcode`@=12
\begin{document}
\thispagestyle{empty}
\begin{flushright}
UCRHEP-T494\\
October 2010\
\end{flushright}
\vspace{0.3in}
\begin{center}
{\Large \bf Neutral SU(2) Gauge Extension of the Standard Model\\
and a Vector-Boson Dark-Matter Candidate\\}
\vspace{1.5in}
{\bf J. Lorenzo Diaz-Cruz$^1$ and Ernest Ma$^2$\\}
\vspace{0.2in}
{\sl $^1$ Facultad de Ciencias Fisico-Matematicas, \\
Benemerita Universidad Autonoma de Puebla, Puebla, Mexico \\
$^2$ Department of Physics and Astronomy, University of California,\\ 
Riverside, California 92521, USA\\}
\end{center}
\vspace{1.2in}
\begin{abstract}
If the standard model of particle interactions is extended to include 
a neutral $SU(2)_N$ gauge factor, with $SU(3)_C \times SU(2)_L \times 
U(1)_Y \times SU(2)_N$ embedded in $E_6$ or $[SU(3)]^3$, a conserved 
generalized $R$ parity may appear.  As a result, apart from the recent 
postulate of a separate non-Abelian gauge factor in the 
hidden sector, we have the first example of a possible dark-matter 
candidate $X_1$ which is a non-Abelian vector boson coming from a 
{\it known} unified model. Using current data, its mass is predicted to 
be less than about 1 TeV.  The associated $Z'$ of this model, as well as 
some signatures of the Higgs sector, should then be observable at the LHC 
(Large Hadron Collider).
\end{abstract}

%\end{document}
\newpage
\baselineskip 24pt
\noindent \underline{\it Introduction}~:~  
Whereas dark matter~\cite{bhs05} is generally accepted as being an important 
component of the Universe, its nature remains unclear.  Myriad hypotheses 
exist, but so far, almost all particles which have been considered as 
dark-matter candidates are spin-zero scalars, or spin-one-half fermions, or 
a combination of both~\cite{cmwy07}.  Spin-one Abelian vector bosons are also 
possible, but only in the context of more exotic scenarios, such as those of 
universal extra dimensions~\cite{LKPdm} and little Higgs models~\cite{LHPdm}.
Spin-one non-Abelian vector bosons from a hidden sector have also been 
considered~\cite{h09,ht10,ahiw10}.  In this paper, we will show for the 
first time that a spin-one non-Abelian vector boson which interacts 
directly with known quarks and leptons may also be a dark-matter candidate, 
motivated by an extension of the standard $SU(3)_C \times SU(2)_L \times 
U(1)_Y$ gauge model of particle interactions with an extra neutral $SU(2)_N$ 
gauge factor, which is derivable from a decomposition of $E_6$ or $[SU(3)]^3$.

We will show how a conserved generalized lepton number may be defined, in 
analogy with the previously proposed dark left-right gauge 
models~\cite{klm09,m09,adhm10,klm10,m10}.  The difference is that the vector 
bosons corresponding to $W_R^\pm$ are now {\it electrically neutral} and 
may become dark-matter candidates.  We will also show how the decomposition 
of $E_6$ or $[SU(3)]^3$ leads to three different models of the form 
$SU(3)_C \times SU(2)_L \times SU(2)' \times U(1)'$.  The first is the 
conventional left-right model where $SU(2)' = SU(2)_R$ and $U(1)' = 
U(1)_{B-L}$, the second is the alternative left-right model~\cite{m87}, 
and the third is the case~\cite{lr86} where $U(1)' = U(1)_Y$ and $SU(2)' 
= SU(2)_N$, with some of its $Z'$ phenomenology already 
discussed~\cite{Nie:2001ti}.  We do not use the original notation of 
$SU(2)_I$, where the subscript $I$ stands for ``inert'', because this 
new gauge group certainly has {\it interactions} linking the known quarks 
and leptons with the exotic fermions.

We will discuss the phenomenology of this model, assuming that the real 
vector gauge boson $X_1$ of $SU(2)_N$ is the lightest particle of odd $R$ 
parity, where $R = (-1)^{3B+L+2j}$, to account for the dark-matter relic 
abundance of the Universe.  This is a new and important possibility not 
discussed previously in the applications of this model.  Combining it with 
the recent CDMS data~\cite{cdms10}, we find $m_X$ to be less than about 1 TeV. 
This means that the associated $Z'(=X_3)$ boson (with even $R$ parity) 
should not be much heavier, and be observable at the Large Hadron Collider 
(LHC). The Higgs sector of this model also has some salient characteristics, 
with good signatures at the LHC.  Note that our proposal is very different 
from the hidden-sector case, where all three gauge bosons, i.e. $X_{1,2,3}$, 
would all be dark-matter candidates having the same mass.

\noindent \underline{\it Model}~:~
Under $SU(3)_C \times SU(2)_L \times U(1)_Y \times SU(2)_N$, where $Q = T_{3L} 
+ Y$, the fermion content of this nonsupersymmetric model is given by
\begin{eqnarray}
&& \pmatrix{u \cr d} \sim (3,2,1/6;1), ~~~ u^c \sim (3^*,1,-2/3;1), \\ 
&& (h^c,d^c) \sim (3^*,1,1/3;2), ~~~ h \sim (3,1,-1/3;1), \\ 
&& \pmatrix{N & \nu \cr E & e} \sim (1,2,-1/2;2), ~~~ \pmatrix{E^c \cr N^c} 
\sim (1,2,1/2;1), \\ 
&& e^c \sim (1,1,1;1), ~~~ (\nu^c,n^c) \sim (1,1,0;2),
\end{eqnarray}
where all fields are left-handed.  The $SU(2)_L$ doublet assignments 
are vertical with $T_{3L} = \pm 1/2$ for the upper (lower) entries. 
The $SU(2)_N$ doublet assignments are horizontal with $T_{3N} = \pm 1/2$ 
for the right (left) entries.  There are three copies of the above to 
accommodate the known three generations of quarks and leptons, together 
with their exotic counterparts.  It is easy to check that all anomalies 
are canceled.

Consider a Higgs sector of one bidoublet and two doublets:
\begin{equation}
\pmatrix{\phi_1^0 & \phi_2^0 \cr \phi_1^- & \phi_2^-} \sim (1,2,-1/2;2), ~~~ 
\pmatrix{\eta^+ \cr \eta^0} \sim (1,2,1/2;1), ~~~ (\chi_1^0,\chi_2^0) \sim 
(1,1,0;2).
\end{equation}
The allowed Yukawa couplings are thus
\begin{eqnarray}
&& (d \phi_1^0 - u \phi_1^-) d^c - (d \phi_2^0 - u \phi_2^-) h^c, ~~~ 
(u \eta^0 - d \eta^+) u^c, ~~~ (h^c \chi_2^0 - d^c \chi_1^0) h, \\ 
&& (N \phi_2^- - \nu \phi_1^- - E \phi_2^0 + e \phi_1^0) e^c, ~~~ 
(E \eta^+ - N \eta^0) n^c - (e \eta^+ - \nu \eta^0) \nu^c, \\ 
&& (E E^c - N N^c) \chi_2^0 - (e E^c - \nu N^c) \chi_1^0,
\end{eqnarray}
as well as
\begin{equation}
(E E^c - N N^c) \bar{\chi}_1^0 + (e E^c - \nu N^c) \bar{\chi}_2^0.
\end{equation}
If Eq.~(9) is disallowed, then a generalized lepton number may be defined, 
with the assignments
\begin{equation}
L=0:~u,d,N,E,\phi_1,\eta,\chi_2^0,n^c, ~~~ L=1:~\nu,e,h,\phi_2, ~~~ 
L=-1:~\chi_1^0,
\end{equation}
so that the neutral vector gauge boson $X$ linking $E$ to $e$ has $L=1$. 
In this scenario, $\phi_2^0$ and $\chi_1^0$ cannot have vacuum expectation 
values.  Fermion masses are obtained from the other neutral scalar fields 
as follows: $m_d,m_e$ from $\langle \phi_1^0 \rangle = v_1$; $m_u,m_\nu$ from 
$\langle \eta^0 \rangle = v_3$; $m_h,m_E,m_N$ from $\langle \chi_2^0 \rangle 
= u_2$.  Actually, because of the $Nn^c$ mass term from $v_3$, $N$ pairs up 
with a linear combination of $N^c$ and $n^c$ to form a Dirac 
fermion, leaving the orthogonal combination massless.  We will return to 
the resolution of this problem in a later section.

To forbid Eq.~(9), an additional global U(1) symmetry $S$ is imposed, as 
discussed in the two original dark left-right models~\cite{klm09,klm10}, 
where $S = L \pm T_{3R}$.  Here we have $S = L - T_{3N}$ instead.  An 
alternative solution is to make the 
model supersymmetric, in which case Eq.~(9) is also forbidden.  We note 
that the structure of this model guarantees the absence of flavor-changing 
neutral currents, allowing thus $SU(2)_N$ to be broken at the relatively 
low scale of 1 TeV. 

\noindent \underline{\it $E_6$ origin}~:~
As listed in Eqs.~(1) to (4), there are 27 chiral fermion fields per 
generation in this model.  This number is not an accident, because 
it comes from the fundamental representation of $E_6$ or $[SU(3)]^3 = 
SU(3)_C \times SU(3)_L \times SU(3)_R$.  Under the latter which is the 
maximal subgroup of the former, these fields transform as $(3,3^*,1) + 
(1,3,3^*) + (3^*,1,3)$, i.e.
\begin{equation}
\pmatrix{d & u & h \cr d & u & h \cr d & u & h} + \pmatrix{N & E^c & \nu \cr 
E & N^c & e \cr \nu^c & e^c & n^c} + \pmatrix{d^c & d^c & d^c \cr u^c & u^c & 
u^c \cr h^c & h^c & h^c}.
\end{equation}
The decomposition of $SU(3)_L \to SU(2)_L \times U(1)_{Y_L}$ is completely 
fixed because of the standard model.  However, the decomposition of 
$SU(3)_R \to SU(2)' \times U(1)'$ is not.  If we choose the conventional 
path, then we see from the above that $(\nu^c,e^c)$ and $(u^c,d^c)$ are 
$SU(2)_R$ doublets.  However, another choice is to switch the first and 
third columns of $(1,3,3^*)$ and the first and third rows of $(3^*,1,3)$, i.e.
\begin{equation}
\pmatrix{d & u & h \cr d & u & h \cr d & u & h} + \pmatrix{\nu & E^c & N \cr 
e & N^c & E \cr n^c & e^c & \nu^c} + \pmatrix{h^c & h^c & h^c \cr u^c & u^c & 
u^c \cr d^c & d^c & d^c}.
\end{equation}
This is the alternative left-right model~\cite{m87}, where $(n^c,e^c)$ and 
$(u^c,h^c)$ are $SU(2)_R$ doublets.

The third choice~\cite{lr86} is to switch the second and third columns 
of $(1,3,3^*)$ and the second and third rows of $(3^*,1,3)$, i.e.
\begin{equation}
\pmatrix{d & u & h \cr d & u & h \cr d & u & h} + \pmatrix{N & \nu & E^c \cr 
E & e & N^c \cr \nu^c & n^c & e^c} + \pmatrix{d^c & d^c & d^c \cr h^c & h^c & 
h^c \cr u^c & u^c & u^c}.
\end{equation}
This then results in Eqs.~(1) to (4).

In analyzing $Z'$ models from $E_6$, the usual convention is to define 
the two possible extra $U(1)$ gauge factors as coming from 
$E_6 \to SO(10) \times U(1)_\psi$ and $SO(10) \to SU(5) \times U(1)_\chi$. 
The special case $U(1)_\eta = \sqrt{3/8} ~U(1)_\chi - \sqrt{5/8} ~U(1)_\psi$ 
is often also considered.  Here the $Z'$ of $SU(2)_N$ couples to the 
orthogonal combination, i.e. $\sqrt{5/8} ~U(1)_\chi + \sqrt{3/8} ~U(1)_\psi$.  
Under the conventional $SU(3)_R$ assignments, this is equivalent to 
$(1/2)T_{3R} - (3/2)Y_R$, hence $n^c$ is +1/2 and $\nu^c$ is $-1/2$ as 
expected.

\noindent \underline{\it Gauge boson masses}~:~
The extra gauge symmetry $SU(2)_N$ is completely broken by $\langle \chi_2^0 
\rangle = u_2$, so that each of the three gauge bosons $X_{1,2,3}$ has the 
same mass, i.e. $m_X^2 = (1/2) g_N^2 u_2^2$.  Whereas $X_3$ should be 
identified with the extra $Z'$ of this model, coupling to fermions 
according to $T_{3N}$, $(X_1 \mp i X_2)/\sqrt{2}$ are the neutral analogs 
of $W_R^\pm$ with $L = \pm 1$.

With the Higgs content of Eq.~(5), there is a massless fermion per generation, 
corresponding to a linear combination of $n^c$ and $N^c$.  At the same time, 
the neutrino has only a Dirac mass, from the pairing of $\nu$ with $\nu^c$. 
Consider then the addition of the scalar triplet
\begin{equation}
(\xi_3^0,\xi_4^0,\xi_5^0) \sim (1,1,0;3),
\end{equation}
with $S=1$, so that $\xi_3^0$ couples to $n^c n^c$ and $\xi_5^0$ couples to 
$\nu^c \nu^c$.  Let these have nonzero vacuum expectation values $u_3$ and 
$u_5$ respectively, then $L$ is broken by the latter to $(-1)^L$ so that 
$\nu$ gets a seesaw Majorana mass in the usual way.  There is also a large 
Majorana mass for $n^c$, so that no massless particle remains.  At the 
same time, $R$ parity, i.e. $R = (-1)^{3B+L+2j}$, remains valid.  All 
standard-model particles have even $R$.  New particles of even $R$ are 
$\phi_1,\eta,\chi_2^0,Z'$; those of odd $R$ are $N,E,n^c,h,\phi_2,\chi_1^0,
X_{1,2}$, the lightest of which is stable and a good dark-matter candidate 
if it is also neutral. However, $N$ and $\phi_2^0$ are ruled out by 
direct-search experiments because they have $Z$ interactions; $n^c$ and 
$\chi_1^0$ are also ruled out because they are mass partners of $N$ and 
$\phi_2^0$.  That leaves only $X_{1,2}$.

The masses of the gauge bosons are now given by
\begin{eqnarray}
&& m_W^2 = {1 \over 2} g_2^2 (v_1^2 + v_3^2), ~~~ m_{X_{1,2}}^2 = {1 \over 2} 
g_N^2 [u_2^2 + 2(u_3 \mp u_5)^2], \\ 
&& m^2_{Z,Z'} = {1 \over 2} \pmatrix{(g_1^2+g_2^2)(v_1^2+v_3^2) & -g_N 
\sqrt{g_1^2+g_2^2} v_1^2 \cr -g_N \sqrt{g_1^2+g_2^2} v_1^2 & g_N^2 [u_2^2 + 
v_1^2 + 4(u_3^2+u_5^2)]}.
\end{eqnarray}
Note that $X_{1,2}$ are split in mass, and the splitting is typically 
large.  There is also $Z-Z'$ mixing, which is approximately given by 
$-(\sqrt{g_1^2+g_2^2}/g_N)(v_1^2/[u_2^2 + 4(u_3^2+u_5^2)])$.  Experimentally, 
$m_{Z'}$ is constrained~\cite{Nie:2001ti} to be greater than about 900 GeV, 
and the $Z-Z'$ mixing less than a few times $10^{-4}$.  Here, 
since $v_1$ couples to the $d$ and $e$ sectors, we may set it at around 
10 GeV, which is then consistent with $SU(2)_N$ breaking to be at the TeV 
scale.  Note that this scale is not motivated by supersymmetry, but rather 
by dark-matter phenomenology as detailed below.

% New comment on EWPT

The new particles $h,h^c,\nu^c,n^c$ are singlets with respect to the 
standard-model gauge group, whereas the doublets $(N,E)$ and $(E^c,N^c)$ 
obtain masses through $\langle \chi_2^0 \rangle$, which is a standard-model 
singlet.  This means that their contributions to the oblique electroweak 
parameters are negligible and will not upset the precision tests of the 
standard model.  The $SU(2)_N$ interactions of $d^c$ are crucial for 
dark-matter search and dark-matter relic abundance.  They, as well as 
those of $e^c$, are not in conflict with any known experimental constraint. 
There is also no mixing between the standard-model fermions and the new ones 
of this model, because the former have even $R$ and the latter odd.

\noindent \underline{\it $X_1$ as dark-matter candidate}~:~
Assuming that $X_1$ is the lightest particle of odd $R$, its relic 
abundance is easily estimated.  The nonrelativistic cross section of 
$X_1 X_1$ annihilation to $d \bar{d}$, $\nu \bar{\nu}$, $e^- e^+$, and 
$\phi_1 \phi_1^\dagger$ through $h$, $N$, $E$, and $\phi_2$ exchange 
respectively, multiplied by their relative velocity, is given by
\begin{equation}
\sigma v_{rel} = {g_N^4 m_X^2 \over 72 \pi} \left[ \sum_h {3 \over 
(m_h^2+m_X^2)^2} + \sum_E {2 \over (m_E^2 + m_X^2)^2} + {2 \over 
(m_{\phi_2}^2 + m_X^2)^2} \right],
\end{equation}
where the sum over $h$, $E$ is for 3 generations.  The factor of 3 for $h$ is 
the number of colors, the factor of 2 for $E$ and $\phi_2$ is to include $N$ 
which has the same mass of $E$ and the two $SU(2)_L$ components of $\phi_2$. 
Note that there is no $X_1 X_1 Z'$ interaction; only $X_1 X_2 Z'$ is allowed. 
We take the usual {\it ansatz} that $\sigma v_{rel}$ is about 1 pb to 
account for the dark-matter relic abundance.  Assuming that $g_N^2 = g_2^2 
= e^2/\sin^2 \theta_W \simeq 0.4$, and setting all exotic particle masses 
equal, i.e. $m_h = m_E = m_{\phi_2}$, we find $(m_h^2 + m_X^2)/m_X \simeq 
2.16$ TeV.  Since $m_X < m_h$ must hold in this scenario, an upper bound 
of 1.08 TeV on $m_X$ is obtained.

The interaction of $X_1$ with nuclei is only through the $d$ quark, i.e. 
$X_1 d \to h \to X_1 d$.  The coherent spin-independent elastic cross 
section is given by
\begin{equation}
\sigma_0 = {3 g_N^4 m_r^2 \over 512 \pi (m_h^2-m_X^2)^2} {[Z + 2(A-Z)]^2 \over 
A^2},
\end{equation}
where $m_r$ is the reduced mass which is just the nucleon mass for heavy 
$m_X$, and $(Z,A)$ are the atomic and mass numbers of the target nucleus, 
which we take to be $^{73}$Ge, i.e. $Z=32$ and $A-Z=41$.  The recent 
experimental bound~\cite{cdms10} on $\sigma_0$ is very well approximated 
in the range $0.3 < m_X < 1.0$ TeV by the expression~\cite{klm10}
\begin{equation}
\sigma_0 < 2.2 \times 10^{-7}~{\rm pb} ~(m_X/1~{\rm TeV})^{0.86}.
\end{equation}
Using this, we obtain
\begin{equation}
m_h^2 > m_X^2 + (1.03~{\rm TeV})^2~(1~{\rm TeV}/m_X)^{0.43}.
\end{equation}
Here $m_h$ refers only to the mass of the $h$ leptoquark which couples 
to $d$ of the nucleon.  It is easy to see that this requires $m_h > 1.29$ 
TeV.  Combining Eq.~(20) with $\sigma v_{rel} = 1$ pb from Eq.~(17), the 
prediction of our model is then
\begin{equation}
m_X < 1.03~{\rm TeV}.
\end{equation}
To obtain a lower bound on $m_X$, we assume that no exotic particle has 
a mass greater than $\sqrt{5}$ times $m_X$ (basically from requiring that 
no Yukawa coupling exceeds 1). In that case, Eq.~(20) implies $m_X > 0.58$ 
TeV.  It is also interesting to note that for a real-vector-boson dark-matter 
candidate, a large spin-dependent cross section is allowed~\cite{sdpdm}. 

\begin{figure}[htb]
\begin{center}
\includegraphics[width=0.9\textwidth]{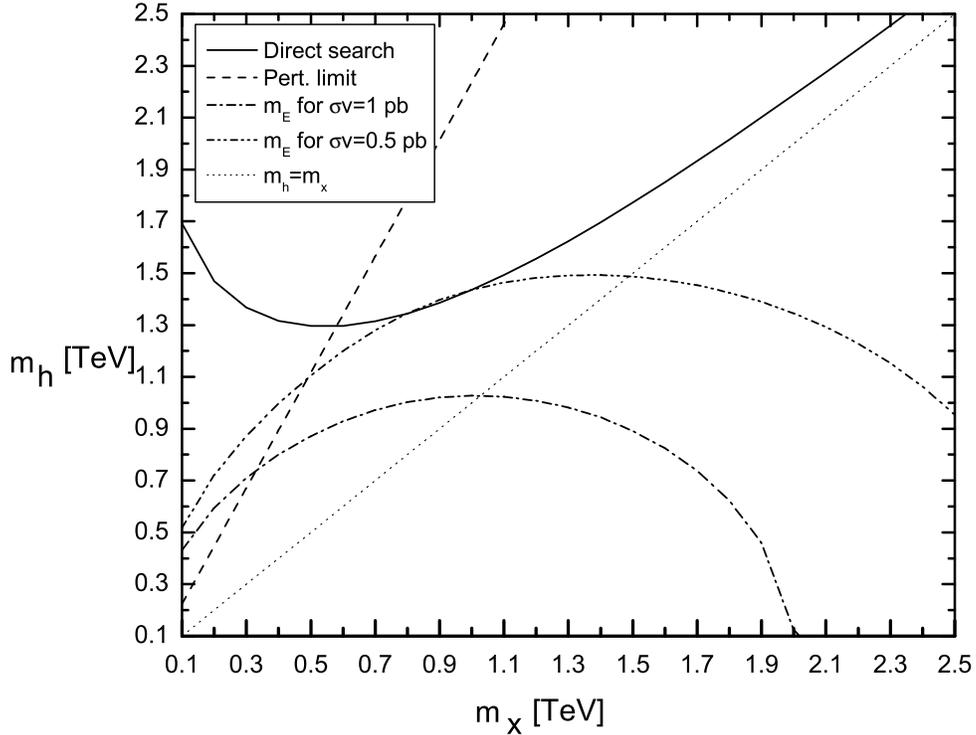}
\caption{Lower bound on $m_h$ versus $m_X$ from CDMS and upper bound on the 
mass of at least one other exotic particle for $\sigma v_{reL} = 1$ pb and 
0.5 pb.}
\end{center}
\end{figure}
In Fig.~1 we show the lower bound of $m_h$ (for the $h$ which couples to 
the $d$ quark of the nucleon) versus $m_X$ from Eq.~(20), as well as the upper 
bound on the mass of at least one other exotic particle (call it $m_E$), 
corresponding to 1 pb and 0.5 pb for $\sigma v_{rel}$ of Eq.~(17).  We impose 
the constraint that $m_X$ is lighter than all other exotic particles and 
assume that no mass is greater than $\sqrt{5} m_X$.  We note that 
$\sigma v_{rel} = 0.5$ pb allows $m_X$ to extend to about 1.5 TeV.

\noindent \underline{\it LHC phenomenology}~:~
As mentioned already, the $Z'$ of our model is a particular linear 
combination of the $Z_\psi$ and $Z_\chi$ of $E_6$ models, which have been 
studied widely in the literature.  Its mass is constrained by present data 
to greater than about 1 TeV.  In our model, from Eqs.~(15) and (16), 
$m_{Z'}$ exceeds $m_X$ by the contributions of $u_3$ and $u_5$, 
which should be of oder 1 TeV.  It will decay into all three generations of 
$d \bar{d}$, $\nu \bar{\nu}$, and $e^- e^+$, and into $\phi_1 \phi_1^\dagger$. 
Its decay into $\mu^- \mu^+$ (with branching fraction 1/16) should be 
a very good signature of its observation~\cite{klm09,klm10}.  Once 
$Z'$ is discovered, the ratios $B(Z' \to t \bar{t})/B(Z' \to \mu^- \mu^+) = 0$ 
and $B(Z' \to b \bar{b})/B(Z' \to \mu^- \mu^+) = 3$ should 
discriminate~\cite{gm08} it from other possible $Z'$ models.

The Higgs sector of this model also has some interesting features that could be
tested at the LHC.  For instance, since $m_{d,s,b}$ and $m_{e,\mu,\tau}$ come 
from $v_1$ which is constrained by $Z-Z'$ mixing to be small, $v_1$ itself 
could be of order 10 GeV or less.  This means that a physical neutral Higgs 
boson with a significant component of $\phi_1^0$ will have large Yukawa 
couplings to $b \bar{b}$.  This can induce a large enhancement on the cross 
section for the associated production of some neutral Higgs bosons with 
$b$ quarks, which may be detectable at the LHC.   Writing the Yukawa 
interaction as $y_b \bar{b}b Re(\phi^0_1)$, with  $y_b = \kappa m_b /v$ 
($v=174$ GeV), we can estimate~\cite{bbhlhc} the values of $\kappa$ that 
can be tested at the LHC. In particular, a Higgs mass of (200, 400, 800) 
GeV can be tested with $\kappa \sim (2.7,~4,~8)$, 
which is well within the enhancement that can be achieved in this model for 
$v_1 \sim 10$ GeV. A detailed study of the Higgs sector of this model, 
including these signals, will be given elsewhere.

\noindent \underline{\it Conclusion}~:~
A neutral $SU(2)_N$ gauge extension of the standard model is studied and 
a non-Abelian vector boson $X_1$ is identified as a possible dark-matter 
candidate. This is the first example of such a particle 
coming from a well-motivated unified model, namely $E_6$ or its maximal 
subgroup $[SU(3)]^3 = SU(3)_C \times SU(3)_L \times SU(3)_R$.
We show that the annihilation of $X_1$ to standard-model 
particles through their $SU(2)_N$ interactions may account for the 
dark-matter relic abundance of the Universe.  Together with the constraint 
from the recent CDMS dark-matter direct-search experiment, we find that 
$m_X$ is less than about 1 TeV.  The associated $Z'$ of this model is 
predicted to be not too much heavier and should be observable at the LHC, 
along with some associated Higgs signatures.

\noindent \underline{\it Addendum}~:~
Variants of our model are easily contemplated.  For example, 
the $SU(2)_N$ Higgs triplet $\xi$ may be eliminated, in which case the 
massless state in the $(N,N^c,n^c)$ sector can become massive by 
adding a singlet $n$ with $S=0$.  If a similar singlet $\nu'$ with $S=1$ 
is also added, neutrinos can get a Majorana mass through the inverse seesaw 
mechanism in the $(\nu,\nu^c,\nu')$ sector with a small Majorana mass term 
for $\nu'$ which breaks $L$ to $(-1)^L$ softly.  In this scenario, the 
splitting of $X_{1,2}$ is radiative and finite, but very small. This 
implies a very different phenomenology for dark matter, because $X X^\dagger$ 
annihilation as well as $X d$ scattering through $Z'$ must be taken into 
account.  These and other issues will be discussed elsewhere.

%\newpage
\noindent \underline{\it Acknowledgements}~:~ 
This work was supported in part by the U.~S.~Department of Energy under
Grant No. DE-FG03-94ER40837.  We also acknowledge support from 
UC-MEXUS and CONACYT-SNI (Mexico).

%\newpage
\baselineskip 16pt
\bibliographystyle{unsrt}

\end{document}